%% file: p207.tex
\documentclass[11pt,epsf]{article}
\usepackage{amsmath}
\usepackage{amsfonts}
\usepackage{amssymb}
\usepackage{graphicx}
\usepackage{color}

\newtheorem{theorem}{Theorem}

\newtheorem{definition}{Definition}
\newcommand {\dfn} {\stackrel{\Delta} {=}}
\newcommand {\exe} {\stackrel{\cdot} {=}}
\newcommand {\gexe} {\stackrel{\cdot} {\ge}}
\newcommand {\lexe} {\stackrel{\cdot} {\le}}

\newcommand{\gea}{\stackrel{\mbox{(a)}}{\ge}}
\newcommand{\geb}{\stackrel{\mbox{(b)}}{\ge}}
\newcommand{\gec}{\stackrel{\mbox{(c)}}{\ge}}
\newcommand{\ged}{\stackrel{\mbox{(d)}}{\ge}}
\newcommand{\gee}{\stackrel{\mbox{(e)}}{\ge}}
\newcommand{\gef}{\stackrel{\mbox{(f)}}{\ge}}
\newcommand{\geg}{\stackrel{\mbox{(g)}}{\ge}}
\newcommand {\reals} {{\rm I\!R}}

\newcommand {\bepsilon} {\mbox{\boldmath $\epsilon$}}

\newcommand {\bs} {\mbox{\boldmath $s$}}

\newcommand {\bu} {\mbox{\boldmath $u$}}

\newcommand {\bx} {\mbox{\boldmath $x$}}
\newcommand {\by} {\mbox{\boldmath $y$}}
\newcommand {\bz} {\mbox{\boldmath $z$}}

\newcommand {\bE} {\mbox{\boldmath $E$}}

\newcommand {\bX} {\mbox{\boldmath $X$}}
\newcommand {\bY} {\mbox{\boldmath $Y$}}
\newcommand {\bZ} {\mbox{\boldmath $Z$}}

\newcommand{\calC}{{\cal C}}

\newcommand{\calE}{{\cal E}}

\newcommand{\calH}{{\cal H}}
\newcommand{\calI}{{\cal I}}

\newcommand{\calL}{{\cal L}}

\newcommand{\calO}{{\cal O}}

\newcommand{\calS}{{\cal S}}

\newcommand{\calY}{{\cal Y}}

\allowdisplaybreaks

\begin{document}
\thispagestyle{empty}
\title{Weak--Noise Modulation--Estimation
of Vector Parameters\thanks{This research was supported by the Israel Science
Foundation (ISF), grant no.\ 137/18.}}
\author{Neri Merhav}
\date{}
\maketitle

\begin{center}
The Andrew \& Erna Viterbi Faculty of Electrical Engineering\\
Technion - Israel Institute of Technology \\
Technion City, Haifa 32000, ISRAEL \\
E--mail: {\tt merhav@ee.technion.ac.il}\\
\end{center}
\vspace{1.3\baselineskip}
\setlength{\baselineskip}{1.3\baselineskip}

\begin{abstract}
We address the problem of modulating a parameter onto a power--limited signal,
transmitted over a discrete--time Gaussian channel
and estimating this parameter at the receiver. Continuing an earlier work,
where the optimal trade--off between the weak--noise estimation performance and
the outage probability (threshold--effect breakdown) was studied for a single (scalar)
parameter, here we extend the derivation
of the weak--noise estimation performance to the case of a 
multi--dimensional vector parameter. 
This turns out to be a non--trivial
extension, that provides a few insights and it has some interesting
implications, which are discussed in depth. 
Several modifications and extensions of the basic setup are also
studied and discussed.\\

\end{abstract}

\newpage
\section{Introduction}

We consider the scenario of communicating a real--valued parameter $u$ by
using an additive white Gaussian noise (AWGN) channel $n$ times, that is,
\begin{equation}
y_t=x_t+z_t,~~~~~~~t=1,2,\ldots,n,
\end{equation}
where $x_t$ is the $t$--th coordinate of a channel input vector,
$\bx=(x_1,x_2,\ldots,x_n)=f_n(u)$, which depends on
$u$, and which is limited by a power constraint,
$\|\bx\|^2\le nP$, $\{z_t\}$ are independent, zero--mean, Gaussian
random variables with variance $\sigma^2$, and $y_t$ is the $t$--th component
of the channel output vector, $\by=(y_1,y_2,\ldots,y_n)$.
In the simplest case, the parameter is assumed to 
take on values within a finite interval, say,
$[0,1]$. Generally speaking, our main interest is
in the following question: what is the best one can do in
estimating $u$
from $\by$, given that there is freedom to design, not only the estimator at
the receiver, but
also the modulation function, i.e., the choice of $f_n(\cdot)$?
How rapidly can the estimation error decay as $n$ grows without bound, given
that the optimal modulator and estimator are used?

This problem is actually the discrete--time counterpart of the
classical ``waveform communication'' problem (in the terminology of
\cite[Chap.\ 8]{WJ65}), and it can be approached
either from the information--theoretic perspective or the
estimation--theoretic perspective. Viewed from the information--theoretic
viewpoint, this is a
joint source--channel coding problem (see, e.g., \cite{KGT17} and references
therein), where one source symbol $u$ is conveyed by $n$ channel
uses, and it also belongs to the realm of Shannon--Kotel'nikov
mappings (see, e.g., \cite{Floor08}, \cite{Hekland07}, \cite{HFR09},
\cite{KR07} and
references therein). From the estimation--theoretic perspective, every given
modulator $f_n(\cdot)$ dictates a certain parametric family of conditional densities of
$\by$ given $u$, for which estimation theory provides an arsenal of
lower bounds 
on the estimation performance (most notably, in terms of the mean square error), 
along with good estimators,
e.g., the maximum likelihood (ML) estimator in the non--Bayesian
case, the maximum
a--posteriori (MAP) estimator and the conditional mean, in the Bayesian case,
and many others.

Conceptually, the simplest form of modulation is linear, that is, the case
where $f_n(u)=u\cdot \bs$ for some fixed vector $\bs$ that is independent of
$u$. Here, the ML estimator achieves the 
non--Bayesian Cram\'er--Rao lower bound (CRLB) for unbiased
estimators and the conditional mean estimator is, of course, optimal in the Bayesian
case. However, the class of linear modulators is very limited and much better
performance can be obtained by non--linear modulation, at least in the high
signal--to--noise (SNR) limit. On the other hand, the
inherent caveat of non--linear modulation is the well--known {\it threshold effect}
(see, e.g., \cite[Chap.\ 8]{WJ65}). The threshold effect is a phenomenon of an
abrupt passage between two very different types of behavior when the SNR
crosses a certain critical level. For high SNR,
the mean square error (MSE) behaves essentially
as in linear modulation, where it roughly
achieves the CRLB. Below a certain SNR level, however, the
performance breaks down and
the MSE increases very abruptly. As presented in
\cite[Chap.\ 8]{WJ65}, for a given non--linear modulator, we are able to identify a
certain anomalous error event, or {\it outage event}, whose probability becomes
considerably large when the threshold is crossed.

Existing results on non--linear modulation and estimation
include many performance bounds (see, e.g.,
\cite{Burnashev84}, \cite{Burnashev85}, \cite{Cohn70} as well as references
therein), with no distinction between errors associated with relatively weak
noise and anomalous errors. In other words, both kinds of errors are weighted
in the evaluation of the total MSE under equal footing. One might argue,
however, that it would be very reasonable to make a distinction between these 
two types of errors with the rationale that estimation in the event of outage is
actually meaningless. In fact, some analysis in the spirit of such a
separation appears already in \cite[pp.\ 661--674]{WJ65}, but not quite in a very
formal framework.

A more systematic study along the line of separating weak--noise errors
from anomalous errors, appears in \cite[Section IV.A]{KGT17}, where the
communication system design problem was formulated as a constrained optimization
problem of a transmitter and receiver, so as to minimize the conditional MSE
given that no outage event has occurred,
under the constraint
that the outage probability would be kept below a prescribed (small) constant.
In this problem, the definition of the outage event was left as a degree of
freedom to be optimized, in addition to the optimization of the modulator and
the estimator. It was argued in \cite{KGT17}, that the data--processing lower bound is
asymptotically achieved by a simple system that is based on quantization
of the parameter, followed by a good
channel code, where on the receiver side, the
digital message is decoded and then mapped back to the quantized parameter
value. The outage event is then the error event of the channel coding part and 
the weak--noise MSE is just the quantization error.
There seems to be a gap, however, between the achievability and the
converse bound of \cite{KGT17},
because the data--processing converse bound
corresponds to a situation where there is no freedom to
allow an outage event, whereas the weak--noise upper bound therein
suppresses the contribution of outage event (by definition). On the other hand, 
conditioning on the non--outage event might create
conditional dependence between the source and the channel, in which case the
data processing theorem would no longer apply. Therefore, it is not clear
that the data processing inequality is the best tool for the derivation
of converse bounds for this problem.

More recently, in \cite{p202}, the approach of \cite{KGT17} 
was sharpened in two ways: first,
in both the converse bound and the achievability, an outage
event was formally allowed, and its choice was subjected to optimization.
In other words, the
lower bound and the upper bound in \cite{p202} are consistent with each other
as they are associated with the same setup.
Secondly, instead of constraining the outage probability by a small
constant, the outage probability constraint imposed in \cite{p202} 
referred to a given exponentially decaying 
function of $n$, i.e., $e^{-\lambda n}$ for some
given constant $\lambda > 0$. Under this constraint, the
fastest possible decay rate of the MSE,
or more generally, the most rapid decrease of the expectation of
any symmetric, convex function of the estimation error, was sought.
Specifically, in \cite{p202}, upper and lower bounds were derived, 
which coincide in
the high SNR limit within a certain interval of small $\lambda$.
The achievability scheme proposed therein was based on quantization and
channel coding using
lattice codes with Voronoi cells whose shape is
very close to $n$--dimensional spheres \cite[Chap.\ 7]{Zamir14}, for large $n$.
This scheme is asymptotically optimal universally for a wide family of error
cost functions.

This paper is a further development of \cite{p202}.
Referring to the regime where the outage probability is merely
required to tend to zero, not necessarily exponentially in $n$ (i.e., $\lambda > 0$ is
arbitrarily small),
it extends the results of \cite{p202} from the single (scalar)
parameter to a multi--dimensional vector
parameter, modulating a power--limited signal. The reason for focusing on the
case where $\lambda$ is arbitrarily small ($\lambda\to 0$) will be explained in the sequel.

This extension turns out to be rather non--trivial. To understand the reason,
the following background is important:
the weak--noise lower
bound of the one--dimensional parameter \cite{p202} 
depends on the modulator only via the length of the
signal locus curve,
\begin{equation}
L(f_n)=\int_0^1\bigg\|\frac{\partial f_n(u)}{\partial u}\bigg\|\mbox{d}u,
\end{equation}
and then, by deriving a universal upper bound to $L(f_n)$, 
the converse bound is
obtained. It is
then natural to expect, at least at first thought, 
that when it comes to higher dimensions, the role of
the one--dimensional signal locus length in the lower bound, would 
be replaced by the hyper--area/volume of the signal
manifold in the higher dimension. For example, if the transmitted
vector, $f_n(u,v)$, depends on a 
two--dimensional vector parameter, $(u,v)\in[0,1]^2$, it
is appealing to expect, in view of \cite{p202}, that the lower bound would
now depend on 
the surface area of the two--dimensional signal
manifold, which is given by\footnote{The integrand of the expression of $S(f_n)$
can be thought of as the product $\|\partial f_n(u,v)/\partial
u\|\cdot\|\partial f_n(u,v)/\partial v\|\cdot|\sin\theta(u,v)|$, where
$|\sin\theta(u,v)|=\sqrt{1-\cos^2\theta(u,v)}$, $\theta(u,v)$ being
the angle between the vectors $\partial f_n(u,v)/\partial u$
and $\partial f_n(u,v)/\partial v$. This product is equal to the area of the
infinitesimal parallelogram defined by the vectors $\mbox{d}u\cdot
\partial f_n(u,v)/\partial u$ and
$\mbox{d}v\cdot\partial f_n(u,v)/\partial v$ with the angle of $\theta(u,v)$ in
between.}
\begin{equation}
S(f_n)=\int_0^1\int_0^1\sqrt{\bigg|\frac{\partial f_n(u,v)}{\partial u}\bigg\|^2\cdot
\bigg\|\frac{\partial f_n(u,v)}{\partial v}\bigg\|^2-\left<
\frac{\partial f_n(u,v)}{\partial u},\frac{\partial f_n(u,v)}{\partial
v}\right>^2}\mbox{d}u\mbox{d}v,
\end{equation}
and then the converse would be obtained from an upper bound on $S(f_n)$ (see,
e.g., \cite[Section 2.5]{Floor08}).

Unfortunately, we have not been able to obtain lower bounds that depend the
modulator solely
via $S(f_n)$, even in the two--dimensional case, let alone the parallel
quantities in higher dimensions. 

Fortunately enough, however, it turns out that it is
possible to bypass the need to handle such higher dimensional objects. The idea is
to use one--dimensional
paths that ``scan'' (high--resolution grids of) the higher dimensional signal
manifold with a carefully chosen density, 
and then one can basically use the one--dimensional derivations of \cite{p202}.
But different types of scans may yield different lower bounds, and then the
question is what is the best scanning strategy that would yield a tight
(achievable) lower bound. It turns out that it is best to exhaust the
parameter space along
``diagonal'' straight lines. 
For example, in the two dimensional case of $(u,v)$, the idea is to ``scan'' along
straight lines with
a slope of 45 degrees, that is, the family of lines
$u-v=c$ in the $(u,v)$--plane, for all possible values of 
$c$ with the right resolution.
The matching achievability result will be based on simple 
uniform quantization and channel coding, as
in \cite{p202} and some earlier works. Here too, this scheme is asymptotically
optimal universally for a wide class of error cost functions.

More details on both the converse and the achievability 
will be provided, of course, in the sequel, but one of the interesting
conclusions from our results is that 
in optimal parametric modulation, the various components of the
parameter vector share the same resources of channel capacity, and therefore
there is an inherent trade--off among the achievable accuracies of their estimation: if
more estimation accuracy is given to one component of the parameter vector,
then it comes at the
expense of the accuracy that can be obtained in the other components. A similar
behavior was observed also in earlier studies under different regimes and different
performance metrics, such as the large deviations performance \cite{p152},
general moments of the estimation error \cite{p156}, and the mean square error
in a multiple access modulation--estimation scenario \cite{UKM18}. This fact
is not quite trivial if one takes into account 
that in many parameter estimation problems,
there is no conflict and no interaction between the estimation errors achieved
for the various components of the vector parameter. For instance, consider the
case where the ML estimator is
Fisher--efficient in a parametric model whose CRLB matrix is diagonal,
like the example where the desired signal is a linear combination of several
orthogonal basis functions, corrupted by Gaussian white noise, 
and the parameters are the coefficients of this
linear combination. Here, the mean square error of each coefficient is not
affected by the lack of knowledge of the other coefficients. The
above described conclusion of this work tells us then
that, nevertheless, whenever the modulator is also 
subjected to optimization, the resulting
parametric family would exhibit interactions 
and trade--offs among the various parameters.
On a related note, one of the conclusions from our main result is that the
best trade-off is obtained when all components of the vector parameter
contribute essentially evenly to the total error cost.

We also discuss a few extended versions of our basic communication 
model. One of them is modulation--estimation over the dirty--paper channel,
where estimation performance is essentially unaffected by the lack of
channel state information at the receiver. In another model of unknown channel
interference, we discuss how universal decoding techniques can be harnessed
for the purpose of universal estimation. Finally, we demonstrate how
constraints on the modulated signal structure 
(which limit the freedom in the signal design),
may fundamentally affect performance, by showing that 
larger (tighter) lower bounds may apply. This fact has implications, for
example, in multiple access communication.

The outline of the remaining part of the paper is as follows.
In Section \ref{pfa}, we establish notation conventions, define the problem
more formally, and spell out our assumptions. In Section \ref{main}, we
present the main result and discuss its insights, implications and a few
variations. Section \ref{proof} is devoted to the proof of the main theorem.
Finally, in Section \ref{sof}, we summarize our main conclusions 
and mention a few possible further
questions for future research.

\section{Notation, Problem Formulation and Assumptions}
\label{pfa}

\subsection{Notation}

Throughout the paper, random variables will be denoted by capital
letters and specific values they may take will be denoted by the
corresponding lower case letters. Random
vectors and their realizations will be denoted,
respectively, by capital letters and the corresponding lower case letters,
both in the bold face font. 
For example, the random vector $\bX=(X_1,\ldots,X_n)$, ($n$ --
positive integer) may take a specific vector value $\bx=(x_1,\ldots,x_n)$.
The probability of an event $\calE$ will be denoted by $\mbox{Pr}\{\calE\}$,
and the expectation operator will be
denoted by $\bE\{\cdot\}$. 
For two positive sequences $a_n$ and $b_n$, the notation $a_n\exe b_n$ will
stand for equality in the exponential scale, that is,
$\lim_{n\to\infty}\frac{1}{n}\log \frac{a_n}{b_n}=0$. Similarly,
$a_n\lexe b_n$ means that
$\limsup_{n\to\infty}\frac{1}{n}\log \frac{a_n}{b_n}\le 0$, and so on.
The indicator function
of an event $\calE$ will be denoted by $\calI\{E\}$. Finally, the notation $[x]_+$
will stand for $\max\{0,x\}$.

\subsection{Problem Formulation and Assumptions}

The problem formulation is similar to that of \cite{p202}, but with
a few modifications that accommodate the extensions being addressed. We
consider the following communication system model. The transmitter needs to
communicate to the receiver, a given value of a
vector parameter $\bu=(u_1,\ldots,u_d)\in[0,1]^d$, and to
this end, it uses the channel $n$ times, subject to a given power
constraint. The receiver has to estimate $\bu$ from the $n$ noisy channel
outputs. More precisely, given $\bu\in [0,1]^d$, the transmitter outputs
a vector $\bx=f_n(\bu)\in\reals^n$, 
subjected to the power constraint, $\|\bx\|^2\le nP$,
$P$ being the allowed power.
The received vector is
\begin{equation}
\bY=f_n(\bu)+\bZ,
\end{equation}
where $\bZ\in\reals^n$ is a zero--mean Gaussian noise vector with covariance
matrix $\sigma^2\cdot I$, $I$ being the $n\times n$ identity matrix.
The receiver employs an estimator $\hat{\bu}=g_n[\by]$ (with $\by$
denoting a realization of the random vector $\bY$)
of the vector parameter $\bu$, where $g_n:\reals^n\to[0,1]^d$.
For every $u\in[0,1]^d$, let $\calO_n(\bu)\subset\reals^n$ denote an event
defined in the
space of noise vectors, $\{\bz\}$, henceforth called the {\it
outage event} (or
the {\it anomalous error event}) given $\bu$.
In the sequel, we will also use the notation
\begin{equation}
\calY_n(\bu)\dfn\calO_n(\bu)+f_n(\bu)\equiv\{\by=f_n(\bu)+\bz:~\bz\in\calO_n(\bu)\},
\end{equation}
and similarly,
\begin{equation}
\calY_n^{\mbox{\tiny c}}(\bu)\dfn\calO_n^{\mbox{\tiny c}}(\bu)+f_n(\bu).
\end{equation}

The goal is to propose a communication
system, defined by a modulator,
$f_n(\cdot)$, that complies with the power constraint, and a
receiver, $g_n[\cdot]$, along with a family of outage events,
$\calO_n=\{\calO_n(\bu),~\bu\in [0,1]^d\}$, in order to minimize
\begin{equation}
\label{obj}
\sup_{\bu\in[0,1]^d}\bE\left\{\rho(g_n[f_n(\bu)+\bZ]-\bu)\bigg|\bZ\in\calO_n^{\mbox{\tiny
c}}(\bu)\right\}
\end{equation}
under the constraint 
\begin{equation}
\label{constraint}
\sup_{\bu\in[0,1]^d}\mbox{Pr}\{\calO_n(\bu)\}\le \delta_n \to 0,
\end{equation}
where the expectation $\bE\{\cdot\}$ in (\ref{obj}) and the probability
$\mbox{Pr}\{\cdot\}$ in
(\ref{constraint}) are with respect
to (w.r.t.) the randomness of $\bZ$. Here, $\delta_n$ is a positive sequence
that tends to zero (not necessarily exponentially fast as in \cite{p202}),
$\calO_n^{\mbox{\tiny c}}(\bu)$ is the complement of $\calO_n(\bu)$,
and $\rho:[-1,1]^d\to\reals^+$ is referred to as
the {\it error cost function} (ECF), which is assumed to be a weighted
$L_q$ distance function, i.e.,
\begin{equation}
\label{ecf}
\rho(\bepsilon)=\rho(\epsilon_1,\ldots,\epsilon_d)=\sum_{i=1}^d
W_i\cdot|\epsilon_i|^q,
\end{equation}
where power $q\ge 1$ and the weights, $W_i\ge 0$, $i=1,\ldots,d$, are given
constants. The weights reflect the relative importance of good estimation of
the various components of $\bu$. They are constants in the sense 
that they are independent of
$\bepsilon$, but they are allowed to depend on $n$. 
Since we focus on the asymptotic performance in the exponential scale, it
is reasonable to choose $\{W_i\}$ to be exponential functions of $n$, as
otherwise, they their particular choice will have no effect 
on the exponential order of
the results. In particular, we set 
$W_i=e^{-na_i}$, $i=1,2,\ldots,d$, for some real--valued
constants, $a_1,\ldots,a_d$, that are independent of $n$. Thus, eq.\
(\ref{ecf}) now reads
\begin{equation}
\rho(\bepsilon)=\rho(\epsilon_1,\ldots,\epsilon_d)=\sum_{i=1}^d
e^{-na_i}\cdot|\epsilon_i|^q.
\end{equation}
Let $\calC_n$ denote the class of all families of
outage events, $\{\calO_n\}$, that satisfy (\ref{constraint}).

As in \cite{p202}, we will be interested in
characterizing the fastest possible exponential decay rate of (\ref{obj})
subject to (\ref{constraint}), where $\delta_n$ is allowed to decay
arbitrarily slowly (or even remain fixed, independently of $n$). 
As mentioned in the Introduction, this 
is different from \cite{p202}, where $\delta_n$ was taken to
be an exponential function of $n$, i.e., $e^{-\lambda n}$ for a given $\lambda >
0$. Our present setup then corresponds to the limit $\lambda\downarrow 0$ of
\cite{p202}.\footnote{There are two motivations for focusing on the case
$\lambda\downarrow 0$. The first is that it is coherent with the regimes
described in \cite[Chap.\ 8]{WJ65} and in \cite{KGT17}, and the second is 
that it allows full compatibility between the achievable performance and the
converse bound, unlike the case of a general value of $\lambda$, where the gap
was closed only in the high SNR limit and only for a certain range of values
of $\lambda$. Also, for a general positive $\lambda$, in the achievability scheme
of \cite{p202} only a good, capacity--achieving lattice code must have been
used, whereas
here, any capacity--achieving channel code can be used.}

More formally, given sequences of
modulators $F=\{f_n(\cdot)\}_{n\ge 1}$ (all satisfying the power constraint),
estimators $G=\{g_n[\cdot]\}_{n\ge 1}$, and families
of outage events, $\calO=\{\calO_n\}_{n\ge 1}$ (with
$\calO_n\in\calC_n$), let
\begin{equation}
\label{ecfexponent}
\calE(F,G,\calO)=\liminf_{n\to\infty}\left[-\frac{1}{n}\ln\left(
\sup_{\bu\in[0,1]^d}\bE\left\{\rho(g_n[f_n(\bu)+\bZ]-\bu)\bigg|\bZ\in\calO_n^{\mbox{\tiny
c}}(\bu)\right\}\right)\right].
\end{equation}

\begin{definition}
\label{wnece}
We say that $E$ is an {\it achievable weak--noise error cost
exponent} if there exists a sequence of communication systems $(F,G,\calO)$,
such that the following two conditions hold at the same time:
\begin{enumerate}
\item[(a)] $\calE(F,G,\calO)\ge E$, and
\item[(b)] For any $\bu_1,\bu_2\in[0,1]^d$ and any $E^\prime < E$, if
$\rho(\bu_1-\bu_2)\gexe e^{-nE^\prime}$ then
$\calY_n^{\mbox{\tiny c}}(\bu_1)\cap\calY_n^{\mbox{\tiny
c}}(\bu_2)=\emptyset$ for all sufficiently large $n$, uniformly in
$\bu_1,\bu_2\in[0,1]^d$.
\end{enumerate}
\end{definition}

The rationale behind these two conditions is that they formally describe the
distinction between the weak--noise mode (part (a)) and the outage mode
(part (b)). If the weak--noise error cost is essentially no larger than
$e^{-nE}$ (in terms of the exponential order), then two parameter values whose distance 
(in the sense of the distance function $\rho$)
is exponentially strictly larger than $e^{-nE}$, should be far apart also in the signal
domain, so that
they would not be confused as long as the noise is not exceptionally strong.

Our purpose is to derive upper and lower bounds to the largest achievable
weak--noise error cost exponent as a function of
the SNR, $\gamma=P/\sigma^2$, which will be denoted by $E(\gamma)$.
We also denote by $C(\gamma)$ the capacity of the AWGN with SNR $\gamma$, that is,
\begin{equation}
C(\gamma)=\frac{1}{2}\ln(1+\gamma).
\end{equation}
Finally, the following assumptions will be made.
\begin{enumerate}
\item[A.1] The parameters of the ECF satisfy\footnote{Since the l.h.s.\ of eq.
(\ref{assump}) depends only on the parameters of the ECF and the
r.h.s.\ depends only on the SNR, this can also be viewed as an assumption of a
sufficiently high SNR. In fact, since only the ratios between the weights
$\{W_i\}$ matter, this means that only the differences among $\{a_i\}$ are
important, and then it could be convenient to shift them all such that
$\sum_{i=1}^da_i=0$, in which case (\ref{assump}) would simplify to $d\cdot\max_{1\le i
\le d}a_i\le qC(\gamma)$.}
\begin{equation}
\label{assump}
d\cdot\max_{1\le i \le d}a_i-\sum_{j=1}^da_j \le qC(\gamma).
\end{equation}
\item [A.2] Consider any curve in the parameter space, i.e.,
$\calL=\{\bu(t)=[u_1(t),\ldots,u_d(t)]:~0\le t\le 1\}$, where $u_i(\cdot)$ are 
arbitrary continuous functions, and let $\{f_n[\bu(t)],~0\le t\le 1\}$ 
be the image of this
curve in the signal space. For every positive integer $m$, let $0=t_0^m < t_1^m <
t_2^m < \ldots t_m^m < t_{m+1}^m=1$ be a partition of the unit interval. 
It is assumed that if $\max_{1\le\ell\le
m+1}[t_\ell^m-t_{\ell-1}^m]\to 0$ as $m\to\infty$, then
$\sum_{\ell=1}^{m+1}\|f_n[\bu(t_\ell)]-f_n[\bu(t_{\ell-1})]\|$ tends to a
limit\footnote{This limit is the length of the signal locus curve,
$\{f_n[\bu(t)],~0\le t\le 1\}$.} that is independent of the particular sequence of partitions.
\item [A.3] Definition \ref{wnece} applies to the communication 
system $(F,G,\calO)$ for some $E > 0$.
\end{enumerate}

\section{Main Result}
\label{main}

Let us define
\begin{equation}
E(\gamma)\dfn\sup_{\{(R_1,\ldots,R_d):~R_1+\ldots+R_d\le
C(\gamma)\}}\min_{1\le i\le d}[a_i+qR_i].
\end{equation}
It is easy to show that under Assumption A.1, an equivalent expresion for
$E(\gamma)$ is given by
\begin{equation}
\label{simplified}
E(\gamma)=\frac{qC(\gamma)+\sum_{i=1}^da_i}{d}.
\end{equation}
To see why this is true, observe that on the one hand, $E(\gamma)$ is upper
bounded by
\begin{eqnarray}
E(\gamma)&=&\sup_{\{(R_1,\ldots,R_d):~\sum_iR_i\le C(\gamma)\}}\min_{1\le i\le
d}(a_i+qR_i)\nonumber\\
&\le&\sup_{\{(R_1,\ldots,R_d):~\sum_iR_i\le
C(\gamma)\}}
\frac{1}{d}\sum_{i=1}^d(a_i+qR_i)\nonumber\\
&=&\frac{\sum_{i=1}^d a_i+qC(\gamma)}{d},
\end{eqnarray}
but on the other hand, equality is achieved by
the assignment
\begin{equation}
\label{rateassign}
R_i^*=\frac{C(\gamma)}{d}+\frac{1}{q}\left(\frac{1}{d}
\sum_{j=1}^da_j-a_i\right),~~~~~i=1,2,\ldots,d.
\end{equation}
Assumption A.1 ensures that $R_i^*\ge 0$ for all $i=1,\ldots,d$,
a condition that will be required both in the direct part and the converse part of our
main result, which is the following.
\begin{theorem}
\label{thm1}
Consider the setup defined in Section \ref{pfa} along with assumptions
A.1 -- A.3. Then, for every sequence of modulators
(all satisfying the power constraint)
receivers, and families of outage events $\calO$ (with $\calO_n\in\calC_n$),
the largest achievable weak--noise error cost exponent is $E(\gamma)$, given
in (\ref{simplified}).
\end{theorem}

The proof of Theorem \ref{thm1} appears in Section \ref{proof}.
The remaining part of this section is devoted to a discussion concerning
the insights and the implications associated with this theorem.\\

\noindent
{\bf The achievability.}
As is shown in the achievability part of the proof of Theorem \ref{thm1},
$E(\gamma)$ can be achieved using 
the very simple idea of source coding
followed by capacity--achieving channel coding, that is, separate source-- and
channel coding (see also \cite{p202} and references therein, for the use of the
same approach in somewhat different scenarios). Moreover,
the source coding part does not
even require any sophisticated vector quantization, but simply a uniform scalar
quantization at rate $R_i^*$ 
for each component $u_i$ of $\bu$ separately (see also \cite{p152} and
\cite{p156}). Also, this approach 
is universally optimal no matter what the parameters of the error cost
function, $\{a_i\}$ and $q$, may be, as long as the meet Assumption A.1.
The only thing that depends on these parameters
is the optimal assignment of the various quantization rates, $R_1^*,\ldots,R_d^*$.
It is also interesting to note that the
optimal rates are chosen such that all components of the vector parameter
contribute essentially evenly to the total expected weak--noise error cost.\\

\noindent
{\bf The converse.}
Since the achievability
part of Theorem 1, namely, the
inequality $\sup_{F,G,\calO}\calE(F,G,\calO)\ge E(\gamma)$,
is conceptually simple as described above (and with some more detail in the
proof), the deeper and more interesting part of Theorem 1 is
the converse part, which is the
inequality $\sup_{F,G,\calO}\calE(F,G,\calO)\le E(\gamma)$.
This is because it applies, of course,
to any arbitrary modulation--estimation system, not only systems that
are based on separate source-- and channel coding. The converse part also
establishes the fact the problem of modulation and estimation is fundamentally
intimately related to
channel coding and channel capacity.\\

\noindent
{\bf Performance as a function of $d$.}
We observe from (\ref{simplified}) that 
for a fixed value of $\sum_{i=1}^da_i$ (for example, the case where
$\sum_ia_i=0$),
the best achievable weak--noise error cost 
exponent is inversely proportional to the dimension $d$ of the
parameter vector. This behavior was observed also in \cite{p152} and
\cite{p156}, although different performance metrics were considered in those
works. It is also anticipated even by the data processing bound pertaining to the
Bayesian regime, where the components of the parameter vector are $d$ i.i.d.\
random variables, and the resulting lower bound on the distortion depends
(exponentially) on $C(\gamma)/d$. However, the data processing theorem does
not yield a tight lower bound for this problem, because it lower bounds the
total expected error cost, not just the weak--noise expected error cost.\\

\noindent
{\bf Variations on the model.} The observed intimate relationship between
modulation--estimation and channel coding naturally triggers the consideration
of other, more general
communication scenarios, where results from the Shannon theory can easily be
``imported''.\\

\noindent
{\it 1. The dirty--paper channel.}
One example is modulation--estimation across
the dirty--paper coding model \cite{Costa83}, whose capacity is well known
to be the same as 
without interference. In particular, suppose that the
channel model is now given by
\begin{equation}
\label{gifc}
y_t=x_t+s_t+z_t,~~~~~~t=1,2,\ldots,
\end{equation}
where $s_t$ is the $t$--th component of a random interference signal
$\bs=(s_1,\ldots,s_n)$ (with an arbitrary distribution), 
known non--causally to the transmitter
only, $z_t$ is Gaussian noise as before, and $x_t$ is the $t$--th component of
$\bx=f_n(\bu,\bs)$, which is subjected to the power constraint. 
The result of Theorem \ref{thm1} remains intact for this
model. To see why this is true, observe that in the direct part, 
one may apply the same scalar quantization as before, but replace the
ordinary channel code by a capacity--achieving dirty--paper code
\cite{Costa83}. In the
converse part, assume a genie--aided receiver that has access to $\bs$, and then
apply the converse proof of Theorem \ref{thm1} for every given (fixed) $\bs$ (with
the tube--packing argument therein applying to spheres centered at $\bs$,
instead of the origin).\\

\noindent
{\it 2. Universal decoding at the service of universal estimation.}
As another example, consider again the channel (\ref{gifc}), but with two
differences relative to the dirty--paper setting discussed in the previous
paragraph. The first is that the transmitter has no access to $\bs$,
namely, $\bx=f_n(\bu)$ as before, and the second is that now $\bs$ is no
longer a random signal, but an unknown deterministic signal.
Can we still achieve $E(\gamma)$ when
neither the transmitter nor the receiver have access to $\bs$? The answer
turns out to be affirmative at least when $\bs$ is known to have a certain
structure. In particular, let us assume that
\begin{equation}
s_t=\sum_{i=1}^\infty \alpha_i\phi_{i,t},
\end{equation}
where $\{\phi_{i,t}\}$ is a given set of uniformly
bounded basis functions (for example, sine and cosine functions), and the
sequence of unknown coefficients, $\alpha_1,\alpha_2,\ldots$ is absolutely
summable. In estimation--theoretic terms, the parameters
$\alpha_1,\alpha_2,\ldots$ are nuisance parameters, which cannot be controlled, 
as opposed to $\bu$ which is the
desired vector parameter, whose modulation is controlled by the signal design.
Normally, the presence of unknown nuisance parameters causes degradation in
the estimation performance of the desired parameters. Here, however, this is
not the case, at least as far as weak--noise error cost exponents are
concerned. In \cite{merhav93}, a
universal decoder for this channel was proposed and was shown to achieve the
same random coding error exponent as the ML decoder that knows
$\alpha_1,\alpha_2,\ldots$.
A-fortiori, such a
universal decoder achieves the capacity of this channel, $C(\gamma)$.
Therefore, referring to the above--described achievability scheme for
modulation and estimation, that is
based on scalar quantization of each component of $\bu$, followed by channel
coding, if the receiver employs the universal 
decoder of \cite{merhav93}, instead of the
ML decoder, then the system still achieves $E(\gamma)$, in spite of the
lack of knowledge of the (infinitely many) nuisance parameters. This means that
universal decoding results can be harnessed for the purpose of universal
estimation.\\

\noindent
{\it 3. Tighter converse bounds for signals with a given structure.} 
The purpose of our third and last example is to make the point that in the
presence of structural constraints concerning the signal modulation, which
means less freedom in the signal design, tighter converse bounds may sometimes
be available. Consider, for example, the case of a two--dimensional vector
parameter, $\bu=(u_1,u_2)$, and a modulator that is enforced to have the form
\begin{equation}
\label{structure}
f_n(u_1,u_2)=f_{n,1}(u_1)+f_{n,2}(u_2),
\end{equation}
where $f_{n,1}(u_1)$ is limited to have power that does not exceed $P_1$ and
$f_{n,2}(u_2)$ is limited to power $P_2$, and be orthogonal to $f_{n,1}(u_1)$.
An immediate application of this model is the Gaussian multiple access channel (MAC),
where two users wish to convey their parameters, $u_1$ and $u_2$ (see, e.g., 
\cite{UKM18}). Let
$\gamma_1=P_1/\sigma^2$ and
$\gamma_2=P_2/\sigma^2$ be the corresponding SNRs. 
For simplicity, let us consider the case
$a_1=a_2=0$.
One converse bound for the weak--noise error cost 
exponent is obtained by a simple application of
Theorem 1 for $d=2$, regardless of the structure (\ref{structure}). This gives
\begin{equation}
\label{generic}
E(\gamma_1+\gamma_2)=\frac{qC(\gamma_1+\gamma_2)}{2}.
\end{equation}
Another lower bound on the weak--noise expectation 
of $|\epsilon_1|^q+|\epsilon_2|^q$ is
obtained when each one of these two terms is treated individually, as if the
corresponding parameter was the only unknown parameter (like in the case
$d=1$), whereas the other parameter was given (and then the signal associated with it
could have been subtracted from $\by$). This would give
$e^{-nqC(\gamma_1)}+e^{-nqC(\gamma_2)}\exe
e^{-nq\min\{C(\gamma_1),C(\gamma_2)\}}$.
If one of the SNRs is relatively small, then this lower bound is tighter than
the generic bound (\ref{generic}). 
On the other hand, if $\gamma_1=\gamma_2$, the bound (\ref{generic})
is tighter.

\section{Proof of Theorem \ref{thm1}}
\label{proof}

\subsection{Achievability} 

The achievability of $E(\gamma)$ is easily established as
follows. For a
given, arbitrarily small $\epsilon > 0$, let
$(R_1^*,\ldots,R_d^*)$ be as in (\ref{rateassign}), but with $C(\gamma)$ replaced
by $C(\gamma)-\epsilon$. Now, uniformly quantize each component,
$u_i$ of $\bu$ using $M_i=e^{nR_i^*}$ equally spaced quantization points with
spacings of $1/M_i=e^{-nR_i^*}$ in between. Thus,
the total set of quanitzed vectors forms a Cartesian grid of $\prod_{i=1}^d
M_i=e^{n[C(\gamma)-\epsilon]}$ points. Each one of these quantized vectors is
mapped into a codeword of a good channel code at rate $C(\gamma)-\epsilon$, which has a
small maximum error probability, $\delta_n$. If we define the decoding error event
as the outage
event $\calO_n(\bu)$ of the
communication system, then the weak--noise
error cost is induced by the quantization error only.
Since the absolute value of the quantization error, $\Delta_i$, in $u_i$
cannot exceed $e^{-nR_i^*}/2$, then
\begin{equation}
\rho(\Delta_1,\ldots,\Delta_d)\le\sum_{i=1}^de^{-na_i}
\bigg|\frac{e^{-nR_i^*}}{2}\bigg|^q
\exe\exp\left\{-n\min_{1\le i\le d}(a_i+qR_i^*)\right\}=e^{-n[E(\gamma)-q\epsilon/d]}.
\end{equation}
Owing to the arbitrariness of $\epsilon$, we can approach $E(\gamma)$ as
closely as desired. Note that this achievability scheme essentially satisfies
Definition \ref{wnece}, as it both achieves (arbitrarily closely) the weak--noise error
cost of $E(\gamma)$ (which is part (a) of Definition \ref{wnece}), and it sends
distant points in the parameter space to distant points in the signal space
(part (b) of Definition \ref{wnece}): any two points, $\bu_1$ and $\bu_2$, 
whose distance is larger than
the exponential order of $e^{nE^\prime}$, with any $E^\prime < 
E(\gamma)-q\epsilon/d$, must belong to different quantization
cells in the parameter space, and hence be mapped onto different codewords, whose
decoding regions, $\calY_n^{\mbox{\tiny c}}(\bu_1)$ and
$\calY_n^{\mbox{\tiny c}}(\bu_2)$, in turn are disjoint.\\

\subsection{Converse}

\begin{figure}[ht]
\hspace*{3cm}
\input{grid.pstex_t}
\caption{\small Scanning the two--dimensional grid along diagonal straight
lines.}
\label{grid}
\end{figure}
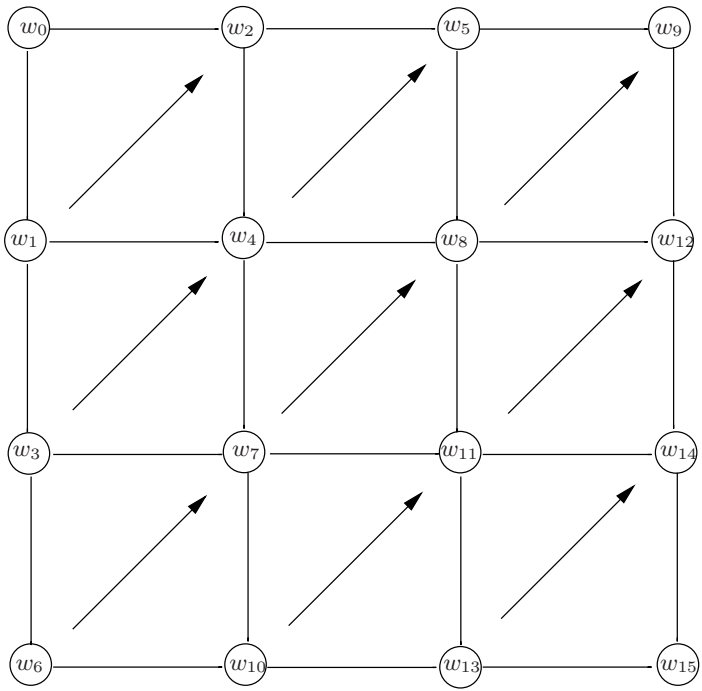

For the sake of simplicity of the exposition, 
we confine attention to the case $d=2$, with the
understanding that the extension to a general dimension $d$ will be
straightforward. The proof is similar to that of the one--dimensional
case of \cite{p202}. The 
difference is that 
instead of referring to a simple path of consecutive grid points along the interval
$[0,1]$, we define a two--dimensional grid and ``scan'' it along diagonal
lines, as demonstrated in Fig.\ \ref{grid}.
To avoid cumbersome notation with many subscripts, let us denote the two
components of the parameter vector by $u$ and $v$, instead of $u_1$ and $u_2$.
Similarly, $a_1$ and $a_2$ will be replaced by $a_u$ and $a_v$, respectively.
For two given positive integers,
$M_u$ and $M_v$ (to be chosen later), consider the uniform grid of points
$(u_i,v_j)\in[0,1]^2$, where $u_i=i/M_u$, $i=0,1,\ldots,M_u-1$, and
$v_j=j/M_v$, $j=0,1,\ldots,M_v-1$. Next, convert this two--dimensional grid
into a one--dimensional sequence, $w_k=(u_i,v_j)$, $k=0,1,\ldots,M_uM_v-1$,
according to the following rule:
for $\ell=(M_v-1),(M_v-2),\ldots,1,0,-1,\ldots,-(M_u-2),-(M_u-1)$,
assign 
\begin{equation}
w_{q(\ell)+i}
=\left(\frac{i}{M_u},\frac{i+\ell}{M_v}\right),~~~~
i=[-\ell]_+,[-\ell]_++1,\ldots,\min\{
M_u,M_v-\ell\}-1,
\end{equation}
where $q(\ell)$ is the number of grid points strictly above the straight line
$v=(u\cdot M_u+\ell)/M_v$ in the $(u,v)$--plane, and the operator $[\cdot]_+$
means positive truncation, i.e., $[t]_+\dfn\max\{0,t\}$.
This conversion into a
one--dimensional sequence is demonstrated in Fig.\ \ref{grid} for $M_u=M_v=4$,
where $u$ and $v$ are designated by the horizontal and vertical axes,
respectively. As can be seen, the scan begins at the top left corner, and then
travels along all parallel diagonal lines (with a slope of 45 degrees), 
one--by--one, each one from the bottom--left to the top--right,
until the entire grid is exhausted.\\

In the derivation below, the function $p(\cdot)$ designates the Gaussian density of the
noise vector $\bz$, that is,
\begin{equation}
p(\bz)=(2\pi\sigma^2)^{-n/2}\exp\left\{-\frac{\|\bz\|^2}{2\sigma^2}\right\}.
\end{equation}
Consider now the following chain of inequalities, which is similar to the one
in \cite{p202}, but with a few twists.
\begin{eqnarray}
\label{longchain}
& &\sup_{(u,v)\in[0,1]^2}\bE\left\{\rho(g_n[\bY]-(u,v))\bigg|\bZ\in\calO_n^{\mbox{\tiny
c}}(u,v)\right\}\nonumber\\
&\ge&\frac{1}{2M_uM_v}\sum_{k=0}^{M_uM_v-2}
\bigg[\bE\bigg(\rho(g_n[\bY]-w_k)\cdot\calI\{\bZ\in\calO_n^{\mbox{\tiny
c}}(w_k)\}\bigg)+\nonumber\\
& &\bE\left(\rho(g_n[\bY]-w_{k+1})\cdot\calI\{\bZ\in\calO_n^{\mbox{\tiny
c}}(w_{k+1})\}\right)\bigg]\nonumber\\
&\gea&\frac{1}{2M_uM_v}\sum_{k\in\calS}
\int_{\calY_n^{\mbox{\tiny c}}(w_k)\cap\calY_n^{\mbox{\tiny c}}(w_{k+1})}
\left[p(\by-f_n(w_k))\rho(g_n[\by]-w_k)+\right.\nonumber\\
& &\left. p(\by-f_n(w_{k+1}))\rho(w_{k+1}-g_n[\by])\right]
\mbox{d}\by\nonumber\\
&\ge&\frac{1}{M_uM_v}\sum_{k\in\calS}
\int_{\calY_n^{\mbox{\tiny c}}(w_k)\cap\calY_n^{\mbox{\tiny c}}(w_{k+1})}
\min\{p(\by-f_n(w_k)),p(\by-f_n(w_{k+1}))\}\times\nonumber\\
& &\left[\frac{1}{2}\rho(g_n[\by]-w_k)+
\frac{1}{2}\rho(w_{k+1}-g_n[\by])\right]
\mbox{d}\by\nonumber\\
&\geb&\frac{1}{M_uM_v}\sum_{k\in\calS}
\int_{\calY_n^{\mbox{\tiny c}}(w_k)\cap\calY_n^{\mbox{\tiny c}}(w_{k+1})}
\min\{p(\by-f_n(w_k)),p(\by-f_n(w_{k+1}))\}\times\nonumber\\
& &\rho\left(\frac{g_n[\by]-w_k}{2}+
\frac{w_{k+1}-g_n[\by]}{2}\right)
\mbox{d}\by\nonumber\\
&=&\frac{1}{M_uM_v}\sum_{k\in\calS}\rho\left(\frac{w_{k+1}-w_k}{2}\right)
\int_{\calY_n^{\mbox{\tiny c}}(w_k)\cap\calY_n^{\mbox{\tiny c}}(w_{k+1})}
\min\{p(\by-f_n(w_k)),p(\by-f_n(w_{k+1}))\}
\mbox{d}\by\nonumber\\
&=&\rho\left(\frac{1}{2M_u},\frac{1}{2M_v}\right)\cdot\frac{1}{M_uM_v}\sum_{k\in\calS}
\int_{\calY_n^{\mbox{\tiny c}}(w_k)\cap\calY_n^{\mbox{\tiny c}}(w_{k+1})}
\min\{p(\by-f_n(w_k)),p(\by-f_n(e_{k+1}))\}\mbox{d}\by\nonumber\\
&=&2\rho\left(\frac{1}{2M_u},\frac{1}{2M_v}\right)\cdot\frac{1}{M_uM_v}\sum_{k\in\calS}
\left[\frac{1}{2}\int_{\reals^n}\min\{p(\by-f_n(w_k)),p(\by-f_n(w_{k+1}))\}\mbox{d}\by
-\right.\nonumber\\
& &\left.-\frac{1}{2}\int_{\calY_n(w_k)\cup\calY_n(w_{k+1})}
\min\{p(\by-f_n(w_k)),p(\by-f_n(w_{k+1}))\}\mbox{d}\by\right]_+\nonumber\\
&\gec&2\rho\left(\frac{1}{2M_u},\frac{1}{2M_v}\right)\cdot\frac{1}{M_uM_v}\sum_{k\in\calS}
\left[Q\left(\frac{\|f_n(w_{k+1})-f_n(w_k)\|}{2\sigma}\right)-\right.\nonumber\\
& &\left.\frac{1}{2}\int_{\calY_n(w_k)}p(\by-f_n(w_k))\mbox{d}\by-
\frac{1}{2}\int_{\calY_n(w_{k+1})}p(\by-f_n(w_{k+1}))\mbox{d}\by\right]_+\nonumber\\
&\ged&2\rho\left(\frac{1}{2M_u},\frac{1}{2M_v}\right)\cdot\frac{1}{M_uM_v}\sum_{k\in\calS}
\left[Q\left(\frac{\|f_n(w_{k+1})-f_n(w_k)\|}
{2\sigma}\right)-\frac{\delta_n}{2}-\frac{\delta_n}{2}
\right]_+\nonumber\\
&=&2\rho\left(\frac{1}{2M_u},\frac{1}{2M_v}\right)\cdot
\frac{|\calS|}{M_uM_v}\cdot\frac{1}{|\calS|}\sum_{k\in\calS}
\left[Q\left(\frac{\|f_n(w_{k+1})-f_n(w_k)\|}{2\sigma}\right)-
\delta_n\right]_+\nonumber\\
&\gee&2\rho\left(\frac{1}{2M_u},\frac{1}{2M_v}\right)\cdot
\frac{(M_uM_v-M_u-M_v)}{M_uM_v}\cdot\left[\frac{1}{|\calS|}\sum_{k\in\calS}
Q\left(\frac{\|f_n(w_{k+1})-f_n(w_k)\|}{2\sigma}\right)-
\delta_n\right]_+\nonumber\\
&\gef&2\rho\left(\frac{1}{2M_u},\frac{1}{2M_v}\right)\cdot
\left(1-\frac{1}{M_u}-\frac{1}{M_v}\right)\times\nonumber\\
& &\left[Q\left(\frac{1}{2\sigma
(M_uM_v-M_u-M_v)}\sum_{k\in S}\|f_n(w_{k+1})-f_n(w_k)\|\right)-\delta_n
\right]_+\nonumber\\
&\geg&2\rho\left(\frac{1}{2M_u},\frac{1}{2M_v}\right)\cdot
\left(1-\frac{1}{M_u}-\frac{1}{M_v}\right)\cdot
\left[Q\left(\frac{\sum_iL([f_n]_i)}
{2\sigma(M_uM_v-M_u-M_v)}\right)-\delta_n\right]_+,
\end{eqnarray}
where $L([f_n]_i)$  
is the total length of the part of the signal locus
curve, obtained as $w=(u,v)$
travels (continuously) along the $i$-th diagonal line, $v=(u\cdot M_u+i)/M_v$.
This length is well defined thanks to Assumption A.2.
The above labeled inequalities are justified as follows:
\begin{enumerate}
\item[(a)] is by the symmetry of the function $\rho$ and
because we are now confining the summation over $\{k\}$ 
to take place across a subset $\calS$ of $\{0,1,\ldots,M_uM_v-2\}$,
which excludes all $M_u+M_v-2$ role--over transitions $w_k\to w_{k+1}$,
i.e., transitions that are associated with
passages from any diagonal line to the next one (in Fig.\ \ref{grid}, the
transitions $w_0\to w_1$, $w_2\to w_3$, $w_5\to w_6$, $w_9\to w_{10}$,
$w_{12}\to w_{13}$ and $w_{14}\to w_{15}$).
\item[(b)] is by the convexity of the function $\rho$, as $q\ge 1$. 
\item[(c)] is by the union bound and by
recognizing the first
integral the line before as the probability of error of the 
ML decision rule in testing the two equiprobable hypotheses:
$\calH_0:~\bY=f_n(w_k)+\bZ$ and
$\calH_1:~\bY=f_n(w_{k+1})+\bZ$, and by relying on the fact that this 
probability of error equal to $Q(\|f_n(w_{k+1})-f_n(w_k)\|/2\sigma)$, where $Q(s)=
\frac{1}{\sqrt{2\pi}}\int_s^\infty e^{-t^2/2}\mbox{d}t$.
\item[(d)] is by the outage constraint (\ref{constraint}).
\item[(e)] is because of the convexity of the function $h(t)=[t-a]_+$ (for
any value of $a$).  
\item[(f)] is due to the convexity of $Q(s)$ for $s\ge 0$ (which can
easily be verified from its second derivative). 
\item[(g)] is because $Q(\cdot)$ is monotonically decreasing and because
$\sum_{k\in S}
\|f_n(w_{k+1})-f_n(w_k)\|\le \sum_iL([f_n]_i)$, which in turn follows from the fact that
the Euclidean distance is a metric, and
so, a straight line between two points is never longer than any
curve in between.
\end{enumerate}
The last step is to upper bound $\sum_iL([f_n]_i)$ by a quantity that is
independent of the particular modulator, $f_n$, and thereby obtain a universal
lower bound to the weak--noise error cost. 

Before we proceed,
we first provide a brief outline of the 
parallel step in \cite{p202}. In the one--dimensional case of
\cite{p202}, there corresponding quantity was $L(f_n)$ (rather than
$\sum_iL([f_n]_i)$), namely, the length of the
entire signal locus, and it was upper bounded by a tube--packing
consideration: the volume of the object
$\{f_n(u)+\bz:~u\in[0,1],~\bz\in\calO_n^{\mbox{\tiny
c}}(u),~\bz\perp\mbox{d}f_n(u)\}$ could not
exceed the volume of the sphere of radius $\sqrt{n(P+\sigma^2)}$, 
of typical channel output signals, but on the other hand, it is lower bounded
by $L(f_n)\cdot\min_u\mbox{Vol}\{\calO_n^{\mbox{\tiny c}}(u)\}$, which in turn
is further lower bounded by $L(f_n)$ times the volume of a sphere of radius
(slightly larger than) $\sqrt{n\sigma^2}$, as a sphere occupies the least
volume among all objects with a given probability (cf.\ (\ref{constraint}) 
under the Gaussian density).
Thus, $L(f_n)$ was upper bounded in \cite{p202} by the ratio between the
volumes of spheres
with radii $\sqrt{n(P+\sigma^2)}$ and $\sqrt{n\sigma^2}$. This ratio is
of the exponential order of $e^{nC(\gamma)}$, 
similarly as in \cite[pp.\ 672--673]{WJ65}
(see also \cite[Subsection 2.2.2]{Floor08}).

Here too, we would like to argue that $\sum_iL([f_n]_i)$ is exponentially
upper bounded by $e^{nC(\gamma)}$. To this end, we need to be convinced that the
spherical tubes surrounding the various parts of the signal locus,
$\{[f_n]_i\}$, are disjoint so that the volume of their union would be
equal to the sum of the volumes. By part (b) of Definition \ref{wnece}, this
in turn will be the case if the parallel diagonal straight lines in the
parameter plane are sufficiently far apart. Specifically, for the system to achieve
a weak--noise error cost exponent of $E(\gamma)+\epsilon$ (with $\epsilon > 0$
being arbitrarily small),
any two points, $w$ and $w^\prime$, along
two different parallel diagonal 
(continuous) straight lines, $v=(u\cdot M_u +i)/M_v$, and
$v=(u\cdot M_u +i^\prime)/M_v$, must be at distance, $\rho(w-w^\prime)$,
exponentially no smaller than $e^{-nE(\gamma)}$.

Let us now choose $M_u=M_u^*\dfn e^{nR_u^*}$ and $M_v=M_v^*\dfn e^{nR_v^*}$ with
\begin{eqnarray} 
R_u^*&\dfn&
\frac{C(\gamma)}{2}+\frac{a_v-a_u}{2q}\\
R_v^*&\dfn& \frac{C(\gamma)}{2}+\frac{a_u-a_v}{2q},
\end{eqnarray}
which are obtained by implementing eq.\
(\ref{rateassign}) for $d=2$.
It can be readily verified by a 
simple distance calculation, that the $\rho$--distance between any two points that
belong to different diagonal lines,
cannot be smaller than
$e^{-nE(\gamma)}\min_{t}[|t|^q+|1-t|^q]=2^{1-q}e^{-nE(\gamma)}$,
which is indeed of the exponential order\footnote{For a general $d$, the
pre-exponential constant factor
is different, but the distance between the closest parallel diagonal lines
continues to be of the exponential order of $e^{-nE(\gamma)}$.}
of $e^{-nE(\gamma)}$. Thus, no matter how small $\epsilon$ may be, 
the tubes surrounding $\{[f_n]_i\}$ must be disjoint
(for large enough $n$), and their total volume cannot exceed the exponential
order of $e^{nC(\gamma)}$.
But since we chose $M_uM_v=M_u^*M_v^*=e^{n(R_u^*+R_v^*)}=e^{nC(\gamma)}$, 
the argument of the $Q$--function in the last line of (\ref{longchain})
tends to a constant, and then
the lower bound in the last line of (\ref{longchain}) becomes of the exponential order of
$\rho(e^{-nR_u^*}/2,e^{-nR_v^*}/2)\exe e^{-nE(\gamma)}$.
It follows that $\sup_{F,G,\calO}E(F,G,\calO)\le E(\gamma)$,
which means that a weak--noise error exponent of $E(\gamma)+\epsilon$ cannot
be achieved no matter how small $\epsilon$ is. Since $\epsilon$ is arbitrarly
small, this completes the proof of Theorem \ref{thm1}.

\section{Summary and Outlook}
\label{sof}

In this paper, we have addressed the problem of modulating a 
parameter onto a power--limited signal
transmitted over a discrete--time Gaussian channel
and estimating this parameter at the receiver. The main contribution
is a non--trivial extension of the main result of \cite{p202},
from a single (scalar) parameter (for zero outage exponent) to a
multi--dimensional vector parameter. The main conclusions from our results are
the following: (i) There is a trade--off among the estimation errors of the
various components of the vector parameter. (ii) On a related note, 
in an optimal modilation--estimation system, all components
of the vector parameter contribute essentially evenly to the total weak--noise estimation
error cost. (iii) The weak--noise error cost exponent is proportional to the
channel capacity and to the power $q$ of the error cost function, and
is inversely proportional to the dimension $d$. (iv) The problem is intimately
related to channel capacity theory, both in the direct part and the converse
part. Finally, several more general communication system models were briefly discussed.

A few other more general models would be interesting to study in future
work. These include the Gaussian colored noise channel model, the presence of
clean or noisy feedback, the interference channel, the broadcast channel and
more.


\end{document}

%% file: grid.pstex_t
\begin{picture}(0,0)%
\includegraphics{grid.pstex}%
\end{picture}%
\setlength{\unitlength}{4144sp}%
\begingroup\makeatletter\ifx\SetFigFont\undefined%
\gdef\SetFigFont#1#2#3#4#5{%
  \reset@font\fontsize{#1}{#2pt}%
  \fontfamily{#3}\fontseries{#4}\fontshape{#5}%
  \selectfont}%
\fi\endgroup%
\begin{picture}(4168,4094)(184,-3521)
\put(4098,-862){\makebox(0,0)[lb]{\smash{{\SetFigFont{9}{10.8}{\rmdefault}{\mddefault}{\itdefault}{$w_{12}$}%
}}}}
\put(4113,-2127){\makebox(0,0)[lb]{\smash{{\SetFigFont{9}{10.8}{\rmdefault}{\mddefault}{\itdefault}{$w_{14}$}%
}}}}
\put(4113,-3406){\makebox(0,0)[lb]{\smash{{\SetFigFont{9}{10.8}{\rmdefault}{\mddefault}{\itdefault}{$w_{15}$}%
}}}}
\put(242,-2127){\makebox(0,0)[lb]{\smash{{\SetFigFont{9}{10.8}{\rmdefault}{\mddefault}{\itdefault}{$w_3$}%
}}}}
\put(258,-3406){\makebox(0,0)[lb]{\smash{{\SetFigFont{9}{10.8}{\rmdefault}{\mddefault}{\itdefault}{$w_6$}%
}}}}
\put(1553,-2127){\makebox(0,0)[lb]{\smash{{\SetFigFont{9}{10.8}{\rmdefault}{\mddefault}{\itdefault}{$w_7$}%
}}}}
\put(4083,401){\makebox(0,0)[lb]{\smash{{\SetFigFont{9}{10.8}{\rmdefault}{\mddefault}{\itdefault}{$w_9$}%
}}}}
\put(1537,-3406){\makebox(0,0)[lb]{\smash{{\SetFigFont{9}{10.8}{\rmdefault}{\mddefault}{\itdefault}{$w_{10}$}%
}}}}
\put(2802,-2127){\makebox(0,0)[lb]{\smash{{\SetFigFont{9}{10.8}{\rmdefault}{\mddefault}{\itdefault}{$w_{11}$}%
}}}}
\put(2817,-3406){\makebox(0,0)[lb]{\smash{{\SetFigFont{9}{10.8}{\rmdefault}{\mddefault}{\itdefault}{$w_{13}$}%
}}}}
\put(284,408){\makebox(0,0)[lb]{\smash{{\SetFigFont{9}{10.8}{\rmdefault}{\mddefault}{\itdefault}{$w_0$}%
}}}}
\put(227,-862){\makebox(0,0)[lb]{\smash{{\SetFigFont{9}{10.8}{\rmdefault}{\mddefault}{\itdefault}{$w_1$}%
}}}}
\put(1537,-847){\makebox(0,0)[lb]{\smash{{\SetFigFont{9}{10.8}{\rmdefault}{\mddefault}{\itdefault}{$w_4$}%
}}}}
\put(2817,417){\makebox(0,0)[lb]{\smash{{\SetFigFont{9}{10.8}{\rmdefault}{\mddefault}{\itdefault}{$w_5$}%
}}}}
\put(1522,401){\makebox(0,0)[lb]{\smash{{\SetFigFont{9}{10.8}{\rmdefault}{\mddefault}{\itdefault}{$w_2$}%
}}}}
\put(2802,-862){\makebox(0,0)[lb]{\smash{{\SetFigFont{9}{10.8}{\rmdefault}{\mddefault}{\itdefault}{$w_8$}%
}}}}
\end{picture}%

%% file: p207.bbl
\clearpage
\begin{thebibliography}{AA}

\bibitem{Burnashev84}
M.~V.~Burnashev, ``A new lower bound for the a-mean error of parameter
transmission over the white Gaussian channel,’’
{\it IEEE Trans.\ Inform.\ Theory}, vol.\ 30, no.\ 1, pp.\ 23--34, January
1984.

\bibitem{Burnashev85}
M.~V.~Burnashev, ``On minimum attainable
mean--square error in transmission of a parameter over a
channel with white Gaussian
noise,’’ {\em Problemy Peredachi Informatsii}, vol.\ 21, no.\ 4, pp.\ 3--16,
1985.

\bibitem{Cohn70}
D.~L.~Cohn, {\it Minimum Mean Square Error Without Coding}, Ph.D.\
dissertation, Massachusetts Institute of Technology, July 1970.

\bibitem{Costa83}
M.~H.~M.~Costa, ``Writing on dirty paper,'' 
{\em IEEE Trans.~Inform.~Theory\/}, vol.\ IT-29, no.\ 5, pp.\ 439--441,
May 1983.

\bibitem{Floor08}
P.~A.~Floor, {\it On the Theory of Shannon-Kotel’nikov Mappings in Joint
Source Channel Coding}, Ph.D.\ dissertation,
Faculty of Information Technology, Mathematics and Electrical Engineering,
Department of Electronics and Telecommunications, Norwegian University of
Science and Technology, May 2008.

\bibitem{Hekland07}
F.~Hekland, {\it On the design and analysis of Shannon-Kotel’nikov Mappings
for Joint Source Channel Coding}, Ph.D.\ dissertation,
Faculty of Information Technology, Mathematics and Electrical Engineering,
Department of Electronics and Telecommunications,
Norwegian University of
Science and Technology, May 2007.

\bibitem{HFR09}
F.~Hekland, P.~A.~Floor, and T.~A.~Ramstad, ``Shannon-Kotel’nikov mappings in
joint source-channel coding,’’ {\it IEEE
Trans.\ Comm.}, vol.\ 57, no.\ 1, pp.\ 94--105, January 2009.

\bibitem{KGT17}
E.~K\"oken, D.~G\"unduz and E.~Tuncel, ``Energy--distortion exponents in
lossy transmission of Gaussian sources over Gaussian channels,’’
{\it IEEE Trans.\ on Inform.\ Theory}, vol.\ 63, no.\ 2, pp.\ 1227--1236,
February 2017.

\bibitem{KR07}
B.~Kravtsov and D.~Raphaeli, ``Multidimensional analog modulation and optimal
mappings,’’ {\it Proc.\ IEEE Globecom 2007}, pp.\ 1426--1430, 2007.

\bibitem{merhav93}
N.\ Merhav, ``Universal decoding for memoryless Gaussian
channels with a deterministic interference,''
{\em IEEE Trans.\ Inform.\ Theory},
vol.~39, no.~4, pp.~1261--1269, July 1993.

\bibitem{p152}
N.~Merhav, ``On optimum
parameter modulation--estimation from a large deviations perspective,''
{\it IEEE Trans.\ Inform.\ Theory}, vol.\ 58, no.\ 12,
pp.\ 7215--7225, December 2012.

\bibitem{p156}
N.~Merhav, ``Exponential error bounds on parameter modulation--estimation for
discrete memoryless channels,''
{\it IEEE Trans.\ Inform.\ Theory}, vol.\ 60, no.\ 2, pp.\ 832--841, February
2014.

\bibitem{p202}
N.~Merhav, ``Trade-offs between weak--noise estimation performance and putage
exponents in nonlinear modulation,’’ to appear in {\it IEEE Trans.\ Inform.\
Theory}, 2019. Also available on--line in {\tt
https://arxiv.org/pdf/1802.04627.pdf}

\bibitem{UKM18}
A.~Unsal, R.~Knopp, and N.~Merhav, ``Converse bounds on modulation--estimation
performance for the Gaussian multiple--access channel,'' {\it IEEE Trans.\
Inform.\ Theory}, vol.\ 64, no.\ 2, pp.\ 1217--1230, February 2018.

\bibitem{WJ65}
J.~M.~Wozencraft and I.~M.~Jacobs, {\it Principles of Communication
Engineering},
John Wiley \& Sons Inc., 1965.

\bibitem{Zamir14}
R.~Zamir, {\it Lattice Coding for Signals and Networks}, Cambridge University
Press, United Kingdom, 2014.
\end{thebibliography}
